\documentclass[%
 reprint,
 amsmath,amssymb,
 aps,
]{revtex4-2}

\usepackage{graphicx}% Include figure files
\usepackage{dcolumn}% Align table columns on decimal point
\usepackage{bm}% bold math
\usepackage{xcolor}
\usepackage{float}
\usepackage{hyperref}% add hypertext capabilities
\begin{document}

\preprint{APS/123-QED}

\title{Probing topological charge of discrete vortices}% Force line breaks with \\
%\thanks{A footnote to the article title}%

\author{Vasu Dev}
\author{Vishwa Pal}%
 \email{vishwa.pal@iitrpr.ac.in}
\affiliation{Department of Physics, Indian Institute of Technology Ropar, Rupnagar, Punjab 140001, India}%

\date{\today}% It is always \today, today,
             %  but any date may be explicitly specified

\begin{abstract}
Discrete vortex, formed by a one-dimensional (1D) ring array of lasers, contains high output power as compared to a conventional continuous vortex, therefore, has attracted considerable interest due to widespread applications in various fields. We present a method for probing the magnitude and sign of the topological charge (TC) of an unknown discrete vortex, by analyzing the interference pattern of a 1D ring array of lasers. The interference pattern of an unknown discrete vortex with TC$\neq0$ is averaged with the interference pattern of TC$=0$, which gives rise to a variation in the fringe visibility as a function of laser number ($j$) in a 1D ring array. The number of dips observed in the fringe visibility curve is found to be proportional to the magnitude of TC of a discrete vortex. The sign of TC is determined by averaging the interference patterns of unknown discrete vortex (TC$\neq0$) with known TC$=+1$. The number of dips in the fringe visibility curve decreases by one for a positive TC, and increases by one for a negative TC. Further, we have verified our method against the phase disorder, and it is found that the phase disorder does not influence an accurate determination of TC of a discrete vortex. The working principle as well as numerical and experimental results are presented for the discrete vortices with TC from small to large values. An excellent agreement between the experimental results and numerical simulations is found. Our method can be useful in the applications of discrete vortices.
\end{abstract}

%\keywords{Suggested keywords}%Use showkeys class option if keyword
                              %display desired
\maketitle

%\tableofcontents

\section{\label{sec:level1}Introduction}
Optical beams with orbital angular momentum (OAM) (also called optical vortex) have attracted considerable attention due to their applications in a wide range of fields- from scientific research to advanced technologies, such as optical communications with increased data carrying capacity \cite{Wang:12, Bozinovic:13}, optical trapping and manipulation \cite{Ng:10, Padgett:11, Tkachenko:14}, microscopy \cite{Hell:94}, quantum information processing \cite{Vaziri:03, Chen:08}, remote sensing \cite{Xie:17}, high-resolution imaging \cite{Maurer:11}, generation of magnetic fields \cite{Lecz:16}, high-harmonic generation \cite{Denoeud:17}, astrophysics for cornography \cite{Berkhout:08}, and probing the angular velocity of spinning microparticles or objects \cite{Martin:13}. An optical vortex possesses a helical wavefront that has a phase singularity in the center, resulting a central dark spot in the transverse field distribution. The optical vortex caries an orbital angular momentum of $l\hbar$ per photon, due to an azimuthal phase term $\exp(il\phi)$, where $l$ is an integer number and is known as the topological charge (TC). Several investigations have been performed on optical vortex beams such as Laguerre-Gaussian (LG) beam \cite{Shen:19}, Bessel-Gaussian beam (BG) \cite{Zhu:08}, and circular Airy vortex beam (CAB) \cite{Liu:17}. These beams consist of continuous intensity around the central dark spot (doughnut shape), and the phase along a contour enclosing the singularity point varies from $0$ to $2l\pi$. Due to continuous intensity and phase over the contour, these beams fall in the category of continuous vortex. The continuous vortex has also been investigated in multimode optical cavities \cite{Arecchi:91}, condensed matter spins \cite{Schliemann:99}, superfluid helium \cite{Zurek:85}, and Bose Einstein condensates (BECs) \cite{Corman:14}.

A new class of discrete vortices have also been investigated in different systems, such as in optical lattices of BECs \cite{Cookson:21}, photonic structure of light \cite{Neshev:04,Malomed:01,Desyatnikov:11,Fleischer:04,Petrovi:09}, coupled lasers \cite{Pal:15,Pal:17,Dev:21}, optical parametric oscillators \cite{Hamerly:16}, and fiber arrays \cite{Alexeyev:09}. Generally, the discrete vortex consists of a finite number of sites (lasers/beamlets/waveguides) in a 1D ring array, where intensity in the center is zero, and the phase circulates from one site to the next either in clockwise (vortex) or anti-clockwise direction (anti-vortex). These consist of step-like behavior of phase along the contour encompassing the phase singularity \cite{Desyatnikov:11,Dev:21,Alexeyev:09}.
The discrete vortices are particularly interesting for high-power applications. These have been generated by different means such as coherent combining of lasers beams \cite{Chu:15,zhi:19,Hou:19,Wang:09,Wang_2009}, where the phases of laser beams are precisely controlled, phase-locking of lasers \cite{Pal:15,Pal:17,Dev:21}, and nonlinear waveguides \cite{Desyatnikov:11}.

The applications of the optical vortex are usually associated with its orbital angular momentum, thus accurately determining the TC values (magnitude and sign) of optical vortex is very important. Therefore, exploring simple and effective methods to accurately measure TC of an optical vortex has been a highly challenging issue, and continuous efforts are growing in this direction. Over the years, several methods have been investigated, which are broadly classified in two categories: first type is based on the interference approach, and the second type is based on the diffraction approach. The information of topological charge is manifested in the intensity distribution when vortex beam either undergoes diffraction through slits/apertures \cite{Ferreira:11, Narag:19}, or interfere with another beam (plane wave/vortex beam/spherical wave) \cite{Lavery:2011, Praveen:22}. Therefore, from the diffraction pattern or interference pattern, the topological charge of vortex beam can be inferred. The interference approaches are mainly based on the Mach-Zehnder interferometer \cite{Lavery:2011, Praveen:22,Ma:21}, Fizeau inteferometer \cite{Cui:19}, Sagnac interferometer \cite{Slussarenko:10}, double-slit interferometer \cite{Sztul:06}, multipoint interferometer \cite{Zhao:20}, and Talbot interferometer \cite{Panthong:18}. Whereas, the diffraction approaches involve the annular aperture \cite{Guo:09}, triangular aperture \cite{deAraujo:11}, single-slit \cite{Ferreira:11, Narag:19}, multi-pinhole plate \cite{Guo1:09}, gratings \cite{Zheng:17}, and metasurfaces \cite{Guo:21}, etc. Further, some more methods for the detection of TC of vortex have also been proposed, which are based on conformal mappings \cite{Wen:18}, multiplane light conversion \cite{Fontaine:19}, mode converter \cite{Zhou:16}, rotational Doppler effect \cite{Zhou:17}, and two-dimensional material \cite{Zhurun:20}. The above methods rely on detecting the topological charges indirectly by analyzing the changes in the intensity distributions. However, some direct methods for determining TC of a vortex have also been investigated. These are based on directly measuring the phase distribution of a vortex, and the method includes phase-shifting digital hologram \cite{Hu:18} and Shack-Hartmann wavefront sensors \cite{Chen:07}. 

Many of these methods pose limitations in various forms, such as complexity in the setup, unable to precisely determine high-order TCs (magnitude and sign), sensitive to aperture dimensions, and sensitive to aberrations in the system. Further, most of these methods infer the TC of a vortex by propagation and analysis of changes in its intensity distribution. These methods have been applied to the continuous systems (continuous vortices). However, it is well-known that discrete systems show different propagation behaviour, so a natural question arises that whether the above methods can be applied or not for identifying the TCs of discrete vortices. 

Methods based on direct measurement of phase distribution of a vortex \cite{Hu:18,Chen:07} can be useful for determining small TC values of discrete vortex, however, given the aberrations as well as very small differences in the phases of lasers for large TCs, these may not be able to precisely determine the high-order TCs of discrete vortices. Here, we present a new method for precisely determining magnitude and sign of TCs (from small to large values) of discrete vortices, based on measuring the interference patterns of 1D ring array of lasers. More specifically, we average the interference pattern of an unknown discrete vortex (TC$\neq 0$) with the interference pattern of ring array with known TC, which gives rise to a variation in the fringe visibility as a function of laser number, and the number of dips observed in fringe visibility provides an accurate information of magnitude and sign of an unknown TC. Our method is also found to be robust against the phase disorder in a system.

The paper is organized as follows. In Sec.\,\ref{principle}, we present the working principle of our method with the illustrative results. In Sec.\,\ref{exp}, we have described the experimental generation of discrete vortex as well as measuring its interference pattern. In Sec.\,\ref{results}, we have presented the numerical and experimental results on precise determination of TCs of discrete vortices for different system sizes. Further, results on finding the sign of TCs (positive/negative) are also presented. The robustness of our method is also verified against the phase disorder. Finally, in Sec.\,\ref{concl}, we present the concluding remarks.
\section{Working principle}\label{principle}
We consider the interference of two waves of intensities $I_{1}$ and $I_{2}$: one propagating in the $z$ direction; the other propagating at an angle $\theta$ with respect to the $z$  axis, in the $x-z$ plane. The fields of these interfering waves can be written as \cite{Saleh07}
\begin{eqnarray}
U_{1}&=&\sqrt{I_{1}} \exp(-ikz)\exp(i\phi_{1}),\label{eq1} \\
U_{2}&=&\sqrt{I_{2}} \exp[-i(k \cos\theta z+k\sin\theta x)] \exp(i\phi_{2}), \label{eq2}
\end{eqnarray}
where $\phi_{1}$ and $\phi_{2}$ are the phases of two waves. 
At the $z=0$ plane, the resultant intensity after superposition of these two waves can be written as
\begin{equation}
%I&=&\left|U_{1}+U_{2}\right|^{2}=|U_{1}|^{2}+|U_{2}|^{2}+U^{*}_{1}U_{2}+U_{1}U^{*}_{2}.
I=\left|U_{1}+U_{2}\right|^{2}=I_{1}+I_{2}+2\sqrt{I_{1}I_{2}} \cos(k\sin\theta x-d\phi), \label{eq3}
\end{equation}
where $d\phi=\phi_{2}-\phi_{1}$. For $I_{1}=I_{2}=I_{0}$, we get
\begin{eqnarray}
I=2I_{0}\left[1+\cos\left(k\sin\theta x-d\phi\right)\right]. \label{eq5} 
\end{eqnarray}
The interference intensity distribution consists of straight fringes with maxima and minima positions vary with $x$, and the fringe spacing is $\beta=2\pi/k\sin\theta=\lambda/\sin\theta$. The locations of minima and maxima can be determined from Eq.\,(\ref{eq5}). 
For maxima:
\begin{equation}
x_{\mathrm{max}}=\frac{[2m\pi+d\phi]}{k\sin\theta}, ~~~m=0,\pm 1,\pm 2,\pm3,... \label{eq7}
\end{equation}
For minima:
\begin{equation}
x_{\mathrm{min}}=\frac{[(2m+1)\pi+d\phi]}{k\sin\theta},~~~ m=0,\pm 1,\pm 2,\pm3,...   \label{eq8}
\end{equation}
The position of the maxima and minima in the interference fringe pattern depends on the value of relative phase shift $d\phi$. By changing it, these maxima and minima positions shift either on the left or right with respect to the initial positions. To show this, we have considered two Gaussian beams with equal intensities $I_{1}=I_{2}=I_{0}$, tilt angle $\theta=110^{o}$ and $d\phi=0 ~\mathrm{and}~ \pm\pi/2$. 
The interference expression for these cases can be written as:\\
(i) For d$\phi=0$
\begin{equation}
 I_{T1}=2I_{0}\left[1+\cos\left(0.94kx\right)\right]. \label{eq9}
\end{equation}
(ii) For d$\phi=-\pi/2$
\begin{equation}
   I_{T2}=2I_{0}\left[1+\cos\left(0.94 kx+\pi/2\right)\right]. \label{eq10}
\end{equation}
(iii) For d$\phi=\pi/2$
\begin{equation}
   I_{T3}=2I_{0}\left[1+\cos\left(0.94 kx-\pi/2\right)\right]. \label{eq11}
\end{equation}
\begin{figure}[htbp]
\centering
\includegraphics[width=8.8cm]{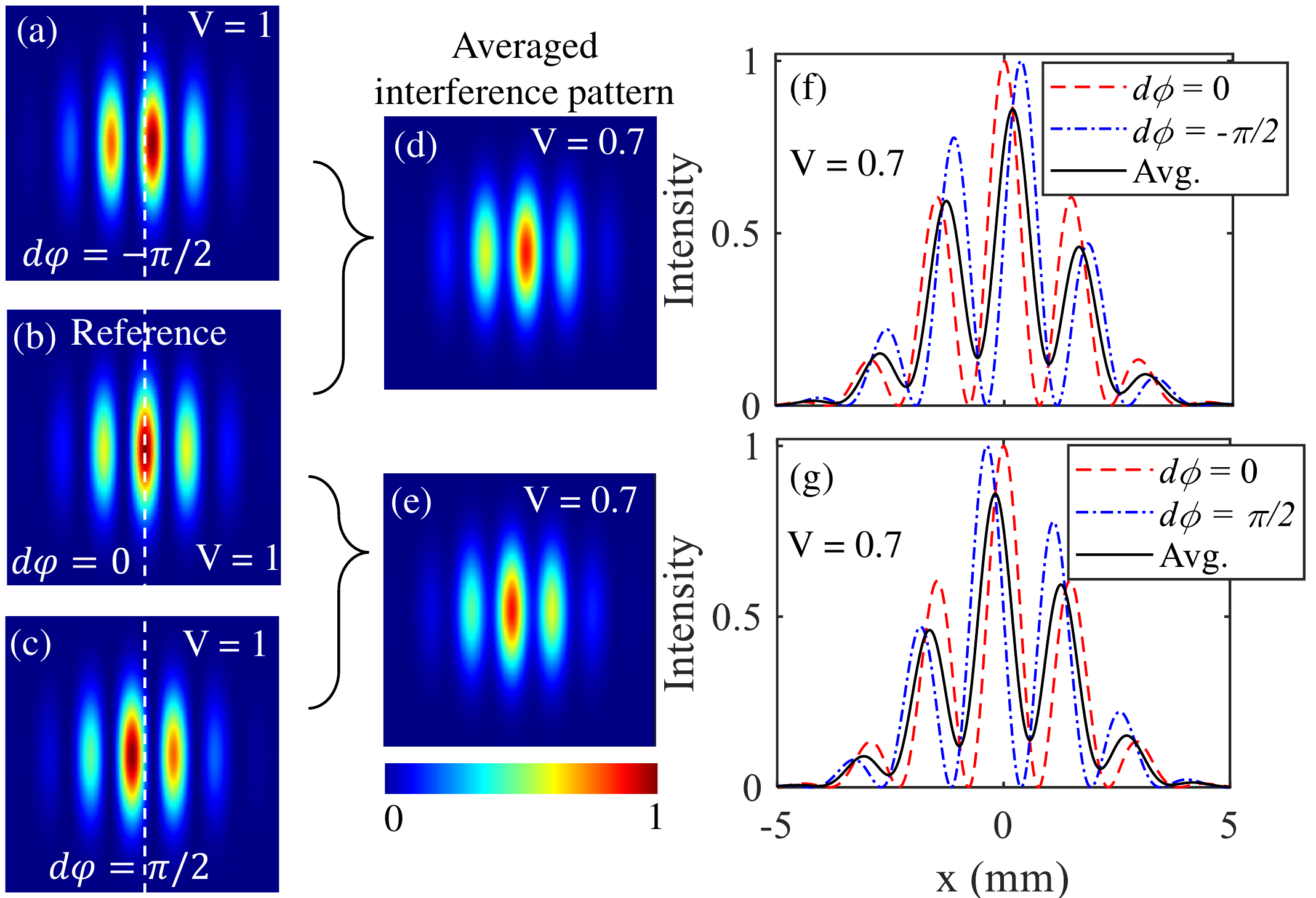}
\caption{\label{fig1} Interference pattern between the two waves having a relative phase shift of (a) $d\phi=-\pi/2$, (b) $d\phi=0$, and (c) $d\phi=\pi/2$. The averaged interference pattern obtained with superposition of (d) $d\phi=0$ and $-\pi/2$, and (e) $d\phi=0$ and $\pi/2$. (f) The intensity cross-sections taken along the horizontal axis in (a), (b) and (d). (g) The intensity cross-sections taken along the horizontal axis in (b), (c) and (e).}
\end{figure}

The interference patterns obtained from Eqs.\,(\ref{eq9})-(\ref{eq11}) are shown in Figs.\,\ref{fig1}(a)-\ref{fig1}(c). As evident, for all three cases, interference patterns consists of straight fringes with maxima and minima at different positions. We consider the case of $d\phi=0$ as the reference interference pattern. A vertical white dashed line marks the position of fringe with central maximum (Fig.\ref{fig1}(b)). For $d\phi=-\pi/2$, the fringes in the interference pattern are shifted towards the right with respect to the reference white dashed line (Fig.\,\ref{fig1}(a)). Similarly, for $d\phi=\pi/2$, the fringes in the interference pattern are shifted towards the left with respect to the reference white dashed line (Fig.\,\ref{fig1}(c)). The interference pattern is quantified by the fringe visibility $V=(I_{\mathrm{max}}-I_{\mathrm{min}})/(I_{\mathrm{max}}+I_{\mathrm{min}})$ \cite{Saleh07}, where $I_{\mathrm{max}}$ and $I_{\mathrm{min}}$ are the maximal and minimal values of the time-averaged intensities in the interference pattern. In all three individual interference patterns (for $d\phi=0,-\pi/2, \mathrm{and}\, \pi/2$) the fringe visibility is found to be $V=1$ (Figs.\,\ref{fig1}(a)-\ref{fig1}(c)). 

Further, we have performed averaging of these interference patterns. The expression of averaged interference pattern can be written as following: (i) For averaging the interference patterns corresponding to Figs.\,\ref{fig1}(a) and \ref{fig1}(b), we add Eqs.(\ref{eq9}) and (\ref{eq10})
\begin{equation}
%    I_{\mathrm{Sum1}}&=&I_{T1}+I_{T2},\label{12}\\
I_{\mathrm{Sum1}}=I_{0}\left[2+\cos\left(0.94kx\right)+\cos\left(0.94 kx+\pi/2\right)\right].\label{eq13}
\end{equation}
Note, due to averaging the factor 2 is dropped.
(ii) For averaging the interference patterns corresponding to Figs.\,\ref{fig1}(b) and \ref{fig1}(c), we add Eqs.(\ref{eq9}) and (\ref{eq11})
\begin{equation}
I_{\mathrm{Sum2}}=I_{0}\left[2+\cos\left(0.94kx\right)+\cos\left(0.94 kx-\pi/2\right)\right].\label{eq14}  
\end{equation}

The averaged interference patterns corresponding to Eqs.\,(\ref{eq13}) and (\ref{eq14}) are shown in Figs.\,\ref{fig1}(d) and \ref{fig1}(e), respectively. As evident, when two interference patterns with $d\phi=0$ and $d\phi=\pm\pi/2$ (each with $V=1$) are averaged, they produce a resultant interference pattern, where the fringe visibility is found to be reduced ($V=0.7$) (Figs.\,\ref{fig1}(d) and \ref{fig1}(e)). 
This is more clearly shown in Figs.\,\ref{fig1}(f) and \ref{fig1}(g). Figure\,\ref{fig1}(f) shows the intensity cross-sections taken along the horizontal axis in Fig.\,\ref{fig1}(a) (blue dot dashed curve), Fig.\,\ref{fig1}(b) (red dashed curve) and Fig.\,\ref{fig1}(d) (black solid curve). Similarly, Fig.\,\ref{fig1}(g) shows the intensity cross-sections taken along the horizontal axis in Fig.\,\ref{fig1}(b) (red dashed curve), Fig.\,\ref{fig1}(c) (blue dot dashed curve) and Fig.\,\ref{fig1}(e) (black solid curve). It is clearly evidenced that the averaged intensity cross-section (black solid curve) results a reduced value of the fringe visibility. Note, averaging the interference patterns corresponding to $d\phi=0$ and $d\phi=\pm\pi$, results the fringe visibility $V=0$.
 
 We utilize this principle of reduction in visibility by averaging the interference patterns, for determining the magnitude and sign of topological charge of an unknown discrete vortex.

\section{Experimental arrangement} \label{exp}
\begin{figure*}[htbp]
\centering
\includegraphics[width=17.0cm]{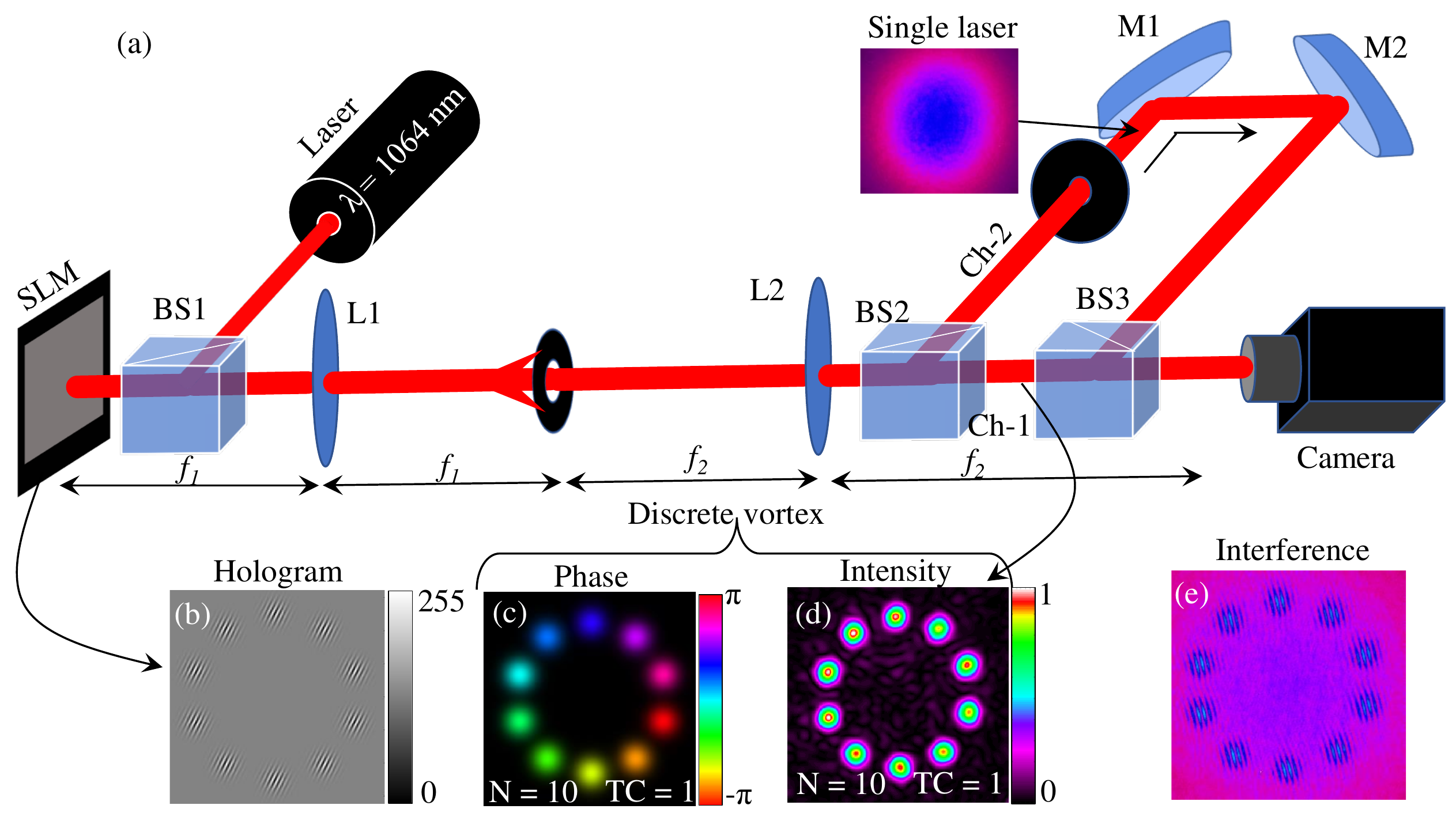}
\caption{\label{fig2} (a) Experimental arrangement for the generation of discrete vortex, and determining its topological charge (including magnitude and sign). SLM: Spatial light modulator; BS1, BS2, BS3: 50:50 Beam splitters; L$_{1}$ and L$_{2}$: Plano-convex lenses with focal lengths $f_{1}$ and $f_{2}$, respectively; M$_{1}$ and M$_{2}$: Mirrors. (b) The phase hologram corresponding to a discrete vortex with TC$=1$. (c) The phase distribution of a discrete vortex with TC$=1$. (d) The intensity distribution of a discrete vortex with N$=10$ lasers. (e) The interference pattern where a single selected reference laser interfered with itself and with all the other lasers.}
\end{figure*}
A discrete vortex consists of a finite number of sites (lasers/beamlets/waveguides) in a 1D ring array, where intensity in the center is zero, and the phase circulates from one site to the next either in clock- wise or anti-clockwise direction \cite{Pal:15,Hou:19,Alexeyev:09, Desyatnikov:11,Hamerly:16}. The field of a discrete vortex can be given as \cite{Wang:09}:
\begin{equation}
    U(x,y;z=0)=U_0 \sum_{j=1}^{N} e^{-\frac{(x-\beta_j)^2 + (y-\gamma_j)^2}{\omega^2_{0}}} e^{i\phi_{j}},
    \label{equ15}
\end{equation}
where $(\beta_{j},\gamma_{j})=\alpha (\cos\delta_{j}, \sin\delta_{j})$ represents the coordinates of each site, $\alpha=d/{\sqrt{1-\cos(2\pi/N)}}$, $\delta_j=\pi (2j-1)/N$, and $\phi_j= \mathrm{TC}.(2\pi (j-1)/N)$, where $j=1,2,3...N$. $d$ denotes the center-to-center separation between two neighboring sites. All sites in a discrete vortex have the same amplitude $U_{0}$, beam waist $\omega_{0}$, and different phase $\phi_{j}$, satisfying the periodic boundary conditions: $U_{j+N} = U_{j}$. The topological charge $TC$ of a discrete vortex can be defined as 
\begin{equation}
    \mathrm{TC}=\frac{1}{2\pi}\sum^{N}_{j=1}\mathrm{arg}(U^{*}_{j}U_{j+1}),
\end{equation}
where $U_{j}$ represents the complex field of site $j$ on a ring. As opposed to a continuous vortex, a discrete vortex is formed by combining several lasers/beamlets/waveguides, thus does not suffer the output power limitations. 

In our previous works, we have shown the generation of discrete vortex by phase-locking lasers in 1D ring array inside a degenerate cavity \cite{Pal:15, Pal:17}. The lasers were coupled by the Talbot diffraction, which caused the phase-locking of lasers in a vortex configuration. However, the discrete vortex with precisely controlled TC could not be generated. More recently, a new approach has been demonstrated for the controlled generation of discrete vortex with any TC from small to large values, where a spatial Fourier filtering mechanism inside a degenerate cavity phase-locks the lasers in a desired vortex configuration \cite{Dev:21}. The spatial Fourier filtering eliminates the undesired phase distributions by introducing extra losses to them, and accordingly only minimum loss solution with desired phase distribution lase inside the cavity.

In the present work, our main goal is to efficiently determine the topological charge (magnitude and sign) of a discrete vortex. To show the proof of concept of our method, we have generated a discrete vortex from a computer generated hologram using a spatial light modulator (SLM). The schematic of experimental setup as well as representative results are presented in Fig.\,\ref{fig2}. We use a phase only SLM with the screen resolution $1920\times 1080$ and pixel size $8~\mu$m. A collimated linearly polarized laser beam with fundamental Gaussian distribution, wavelength $\lambda= 1064$ nm and beam waist radius 10 mm incidents normally on the SLM with a beam splitter (BS1). The size of an input Gaussian beam is chosen such that it illuminates whole screen of SLM. On the SLM screen a computer generated phase hologram corresponding to a discrete vortex is applied (Fig.\,\ref{fig2}(b)) (see Appendix\,\ref{appa}). Phase hologram on the SLM modulates amplitude and phase of an input Gaussian laser beam, and accordingly the light from an input Gaussian laser beam splits in the form of multiple lasers arranged on a 1D ring array with discrete phase distributions in a vortex configuration. After reflection from SLM, we obtain modulated light in several orders (see Appendix\,\ref{appa}), which contains the desired discrete vortex. The desired discrete vortex in the first order is isolated with a spatial Fourier filter placed in the middle of a telescope made with lenses L$_{1}$ ($f_{1}=30$ cm) and L$_{2}$ ($f_{2}=20$ cm). After spatial filtering, we obtain a clean discrete vortex at the focal plane of L$_{2}$. The experimental phase distribution and intensity distribution of a generated discrete vortex are shown in Figs.\,\ref{fig2}(c) and \ref{fig2}(d), respectively. As evident, the discrete vortex consists of N=10 lasers (with TEM$_{00}$ fundamental Gaussian mode profiles) in a 1D ring array (Fig.\,\ref{fig2}(d)), and the lasers possess discrete phase distribution (Fig.\,\ref{fig2}(c)) in a vortex configuration. This is identical to a discrete vortex obtained by phase-locking of lasers in a 1D ring array inside a degenerate cavity \cite{Pal:15, Dev:21}.

To determine the magnitude and sign of topological charge of a discrete vortex, we measure the interference between the lasers using a Mach-Zhender interferometer, as shown in Fig.\,\ref{fig2}(a). The discrete vortex (Figs.\,\ref{fig2}(c) and \ref{fig2}(d)) splits into two channels (Ch-1 and Ch-2) at the beam splitter BS2. In one channel (Ch-1), the discrete vortex is imaged directly onto the camera. In the other channel (Ch-2), a single reference laser is selected with a circular pinhole of suitable size, and then its light is expanded such that it fully overlaps and interferes with the light from all other lasers with a beam splitter BS3 on the camera. This enables that a single selected reference laser interferes with itself and with all the other lasers. A small tilt (Eq.\,(\ref{eq2})) between the two channels provides a few interference fringes for each laser (as shown in the interference pattern Fig.\,\ref{fig2}(e)) from which the fringe visibility is analyzed.

\section{Results and discussions} \label{results}
\begin{figure*}[htbp]
\centering
\includegraphics[width=17.0cm]{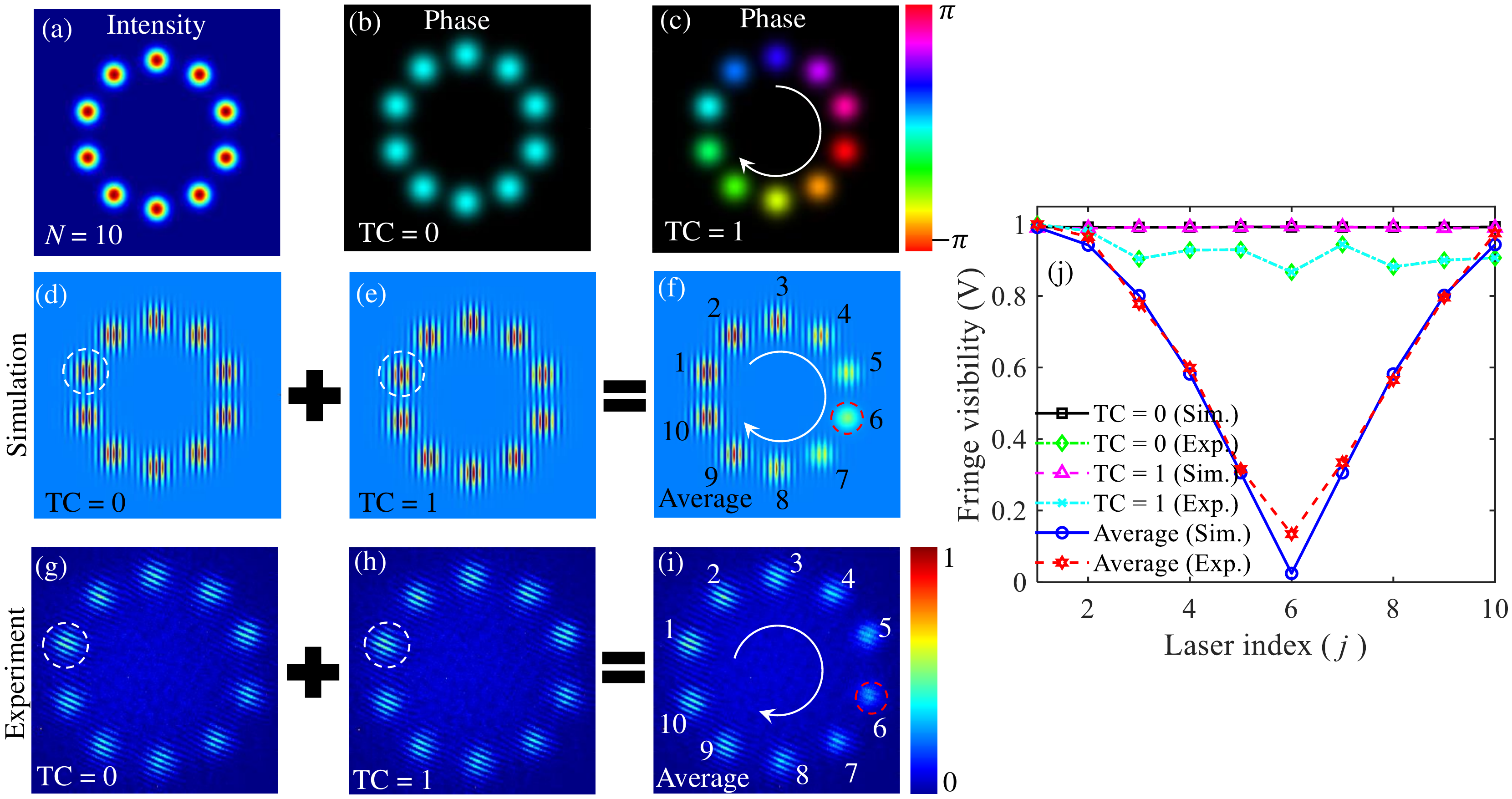}
\caption{\label{fig3} Results for a discrete vortex with TC=1. (a) The near-field intensity distribution of a discrete vortex formed by a 1D ring array of N=10 lasers. The near-field phase distribution with (b) TC=0, and (c) TC=1. Interference pattern when a single laser (reference laser denoted by dashed circle) interferes with itself and with all the other lasers for (d, g) TC=0, and (e, h) TC=1. (f) The average of interference patterns of TC=0 (d) and TC=1 (e). (i) The average of interference patterns of TC=0 (g) and TC=1 (h). (j) The variation of fringe visibility as a function of laser index.}
\end{figure*}

To show that our method can efficiently determine the magnitude and sign of TC (from small to large values) of discrete vortex, we have performed several experiments considering different system sizes N. The experimental results are supported by the numerical simulations. First, we have demonstrated our approach for determining TCs of discrete vortices, formed with a 1D ring array of N=10 lasers. The results for a discrete vortex with TC=1 are shown in Fig.\,\ref{fig3}. 

Figure\,\ref{fig3}(a) shows the intensity distribution of a discrete vortex, indicating that it consists of a 1D ring array of N=10 lasers with TEM$_{00}$ fundamental Gaussian mode profiles, and each laser has the beam waist $0.16$ mm and separated with a distance of $0.84$ mm (center-to-center). Figure\,\ref{fig3}(b) shows the phase distribution of 1D ring array of N=10 lasers with TC=0 (in-phase distribution), indicating that there is no net phase circulation around the center. Figure\,\ref{fig3}(c) shows the discrete phase distribution of 1D ring array of N=10 lasers with TC=1, indicating the phase circulation around the center (discrete vortex). Note, the intensity distribution of 1D ring array of lasers for both TC=0 and TC=1 is the same. Figures\,\ref{fig3}(d)-\ref{fig3}(f) show the simulated results of individual interference patterns of 1D ring array of N=10 lasers with TC=0 and TC=1, and their averaged interference pattern, respectively. Figures\,\ref{fig3}(g)-\ref{fig3}(i) show the experimental results of individual interference patterns of 1D ring array of N=10 lasers with TC=0 and TC=1, and their averaged interference pattern, respectively. The white dashed circles in Figs.\,\ref{fig3}(d), \ref{fig3}(e), \ref{fig3}(g) and \ref{fig3}(h) mark the location of a selected reference laser. The individual interference patterns of 1D ring array of N=10 lasers with TC=0 and TC=1 are obtained using Mach-Zhender interferometer (Fig.\,\ref{fig2}(a)), where a selected reference laser interferes with itself and with all the other lasers. As evident, in the individual interference patterns corresponding to TC=0 and TC=1, the fringes are uniform at all the lasers and show no distinguishabilities between TC=0 and TC=1 (Figs.\,\ref{fig3}(d)\& \ref{fig3}(e) and Figs.\,\ref{fig3}(g) \& \ref{fig3}(h)). However, in the averaged interference pattern the fringes show variation at the lasers (Figs.\,\ref{fig3}(f) and \ref{fig3}(i)). Particularly, at the reference laser $j=1$ the fringes appear with clear maxima and minima, and the fringes degrade monotonically for the lasers $j=2-6$ and after that fringes again improve monotonically for the lasers $j=6-10$. 

We have analyzed the fringe visibility at all the lasers in these individual and averaged interference patterns (see Appendix\,\ref{appb}). The quantified simulated and experimental results are shown in Fig.\,\ref{fig3}(j), which shows the variation of fringe visibility as a function of laser index $j$ ($j=1,~2,~3...10$). As evident, for the individual interference patterns corresponding to TC$=0$ (solid black curve with squares (simulation) and dashed green curve with diamonds (experiment)) and TC$=1$ (pink dashed curve with triangles (simulation) and cyan dashed curve with cross (experiment)), the fringe visibility does not show any variation, and is found to be maximum V$\approx1$ at all the lasers. Note, due to small intensity differences between the lasers, the experimental visibility is found to be a bit less than V$=1$. However, in the averaged interference pattern the fringe visibility shows variation as a function of laser index $j$ (shown by red dashed curve with stars (experiment) and blue solid curve with circles (simulation)). For example, the fringe visibility is found to be maximum (V$=1$) at laser $j=1$ and then it decreases monotonically until laser $j=6$ (minimum $V$), and after that it again increases and becomes maximum again at laser $j=10$. 

These variations in the fringe visibility as a function of laser index can be understood by the working principle given in Sec.\,\ref{principle} (see also Appendix\,\ref{appb}). In case of TC$=0$, all the lasers have zero phase ($\phi_{j}=0$ and $d\phi_{1,j\geq1}=0$, where $j=1,~2,~3...N$), whereas for TC$=1$, the phases of the lasers grow in a clockwise direction ($\phi_{j}=(j-1)2\pi/N$ and $d\phi_{1,j\geq1}=(j-1)2\pi/N$, where $j=1,~2,~3...N$). So the reference laser ($j=1$) in both cases will have the interference fringes with maxima and minima occurring at the same positions, and hence averaging does not reduce the fringe visibility. However, for laser $j\geq2$, we have $d\phi_{1,j\geq2}=0$ in case of TC=0, and $d\phi_{1,j\geq2}=(j-1)2\pi/10$ ($N=10$ in the present case) in case of TC=1. Thus, in the interference patterns of TC=0 and TC=1, the fringes with maxima and minima will occur at different positions in the same laser $j\geq2$ (Eqs.\,(\ref{eq7}) and (\ref{eq8})), and as a result of averaging the fringe visibility is found to be reduced at that laser (Fig.\,\ref{fig1}, and see Appendix\,\ref{appb}). The fringe visibility is found to be lowest at laser $j=6$, where the phase shift $d\phi_{1,6}=\pi$ becomes maximum in TC=1 and the fringes are shifted maximally, thus the averaging results a lowest value of fringe visibility. For lasers $j=7-10$ the fringe visibility again increases monotonically. For laser $j=7$ in TC=1, we have $d\phi_{1,7}=6.(\frac{2\pi}{10})$, which can also be written as [$2\pi-4.(\frac{2\pi}{10})$] or $-4.(\frac{2\pi}{10})$. This becomes equivalent to $-d\phi_{1,5}$. The positive and negative values of the same magnitude $d\phi$ produce the same shift of fringes in the interference pattern, as shown in Fig.\,\ref{fig1}. The magnitude of $d\phi_{1,7}$ is found to be less than $d\phi_{1,6}$, so fringe visibility in the averaged interference pattern at laser $j=7$ is found to be increased, and it is the same as for laser $j=5$. Similarly, the fringe visibility for lasers $j=8,~9 ~\mathrm{and} ~10$ can also be explained (for details, see Appendix\,\ref{appb}). The variation in the fringe visibility as a function of laser index is characteristics to the TC of a discrete vortex, so it serves as the basis for an accurate determination of TC. In particular, the observation of number dips in the fringe visibility curve provides the information of TC. For example, a single minimum denotes the value of TC$=1$. 
\begin{figure*}[htbp]
\centering
\includegraphics[width=14.0cm]{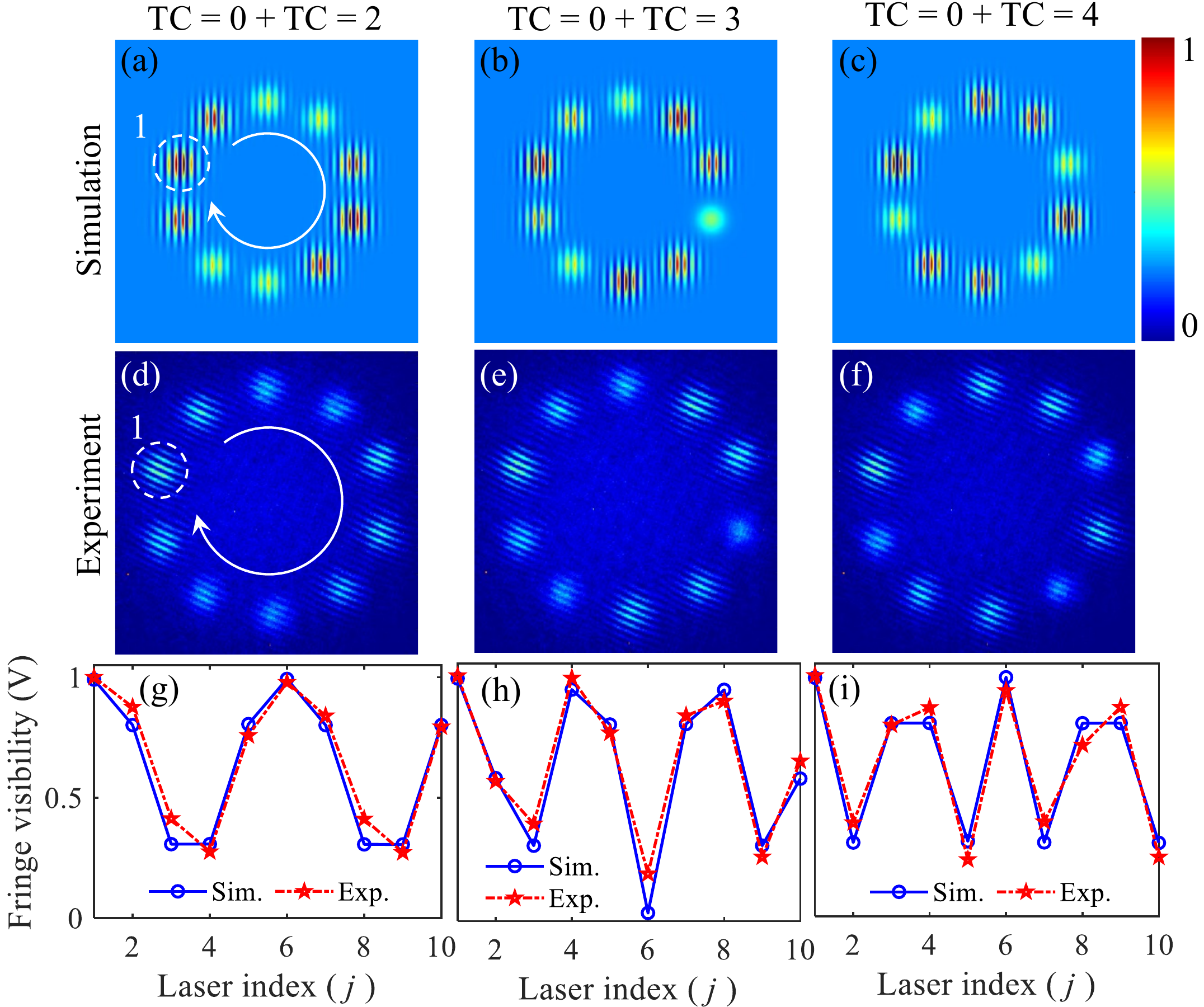}
\caption{\label{fig4} For a discrete vortex of system size N=10 lasers, the determination of TC=2, 3 and 4. The average of interference patterns of TC$=0$ and (a, d) TC$=2$, (b, e) TC$=3$, and (e, f) TC$=4$. (g-i) The Fringe visibility as a function of laser index, corresponding to (a, d), (b, e), and (e, f), respectively. Blue solid curve with circles represents the simulation, and Red dashed curve with stars represents the experiment. Dashed white circle on (a, d) marks the location of a selected reference laser $j=1$.}
\end{figure*}
\begin{figure*}[htbp]
\centering
\includegraphics[width=17.5cm]{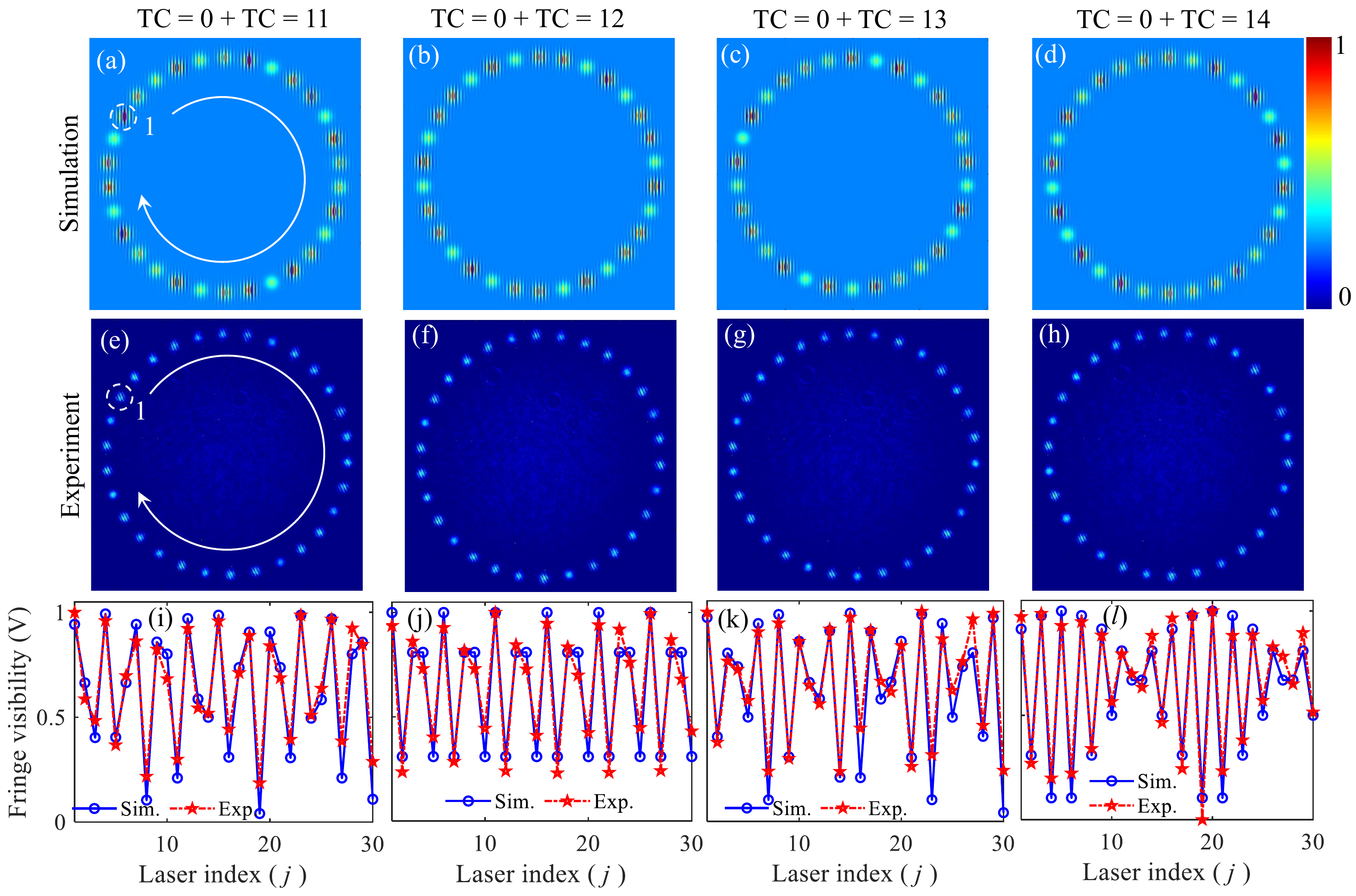}
\caption{\label{fig5} Determination of higher values of TC of a discrete vortex with a large system size N=30 lasers. The average of interference patterns of TC$=0$ and (a, e) TC$=11$, (b, f) TC$=12$, (c, g) TC$=13$, and (d, h) TC$=14$. (e-h)The Fringe visibility as a function of laser index, corresponding to (a, e), (b, f), and (c, g), respectively. Blue solid curve with circles represents the simulation, and red dashed curve with stars represents the experiment. Dashed white circle on (a, e) marks the location of a selected reference laser $j=1$. }
\end{figure*}

Further, we have determined the large values of TC$>1$ in a discrete vortex formed with a 1D ring array of N=10 lasers. The results are shown in Fig.\,\ref{fig4}. Figures\,\ref{fig4}(a)-\ref{fig4}(c) and \ref{fig4}(d)-\ref{fig4}(f) show the simulated and experimental results obtained by averaging the interference patterns of TC=2, TC=3 and TC=4 with TC=0 (TC$=0$ + TC$=2$, TC$=0$ + TC$=3$, and TC$=0$ + TC$=4$), respectively. The white dashed circles in Figs.\,\ref{fig4}(a) and \ref{fig4}(d) mark the location of a selected reference laser $j=1$. The location of reference laser remains the same in other averaged interference patterns (Figs.\,\ref{fig4}(b) $\&$ \ref{fig4}(c) and Figs.\,\ref{fig4}(e) $\&$ \ref{fig4}(f)). As evident, in the averaged interference patterns the fringes at the lasers exhibit different behaviour. The analyzed fringe visibility as a function of laser index ($j$) corresponding to TC$=2$, $3$ and $4$ are presented in Figs.\,\ref{fig4}(g)-\ref{fig4}(i), respectively, indicating different variations related to the values of TCs. This is due to the fact that the phase distribution of discrete vortex depends on the value of TC as $\phi_{j}=\mathrm{TC}.(2\pi(j-1)/10)$ ($j=1,~2,~3...10$), and accordingly the fringes at the lasers are shifted by different amounts with respect to the fringes in TC=0. Thus, averaging of interference patterns of TC$\neq0$ and TC=0 results a different variation in the fringe visibility with the laser index (see Appendix\,\ref{appb}), and can be used to determine the value of TC. For example, the averaging of TC$=0$+TC$=2$ produces two dips in the fringe visibility curve, confirming the value of TC$=2$ (Fig.\,\ref{fig4}(g)). Similarly, the observation of three and four dips in the fringe visibility curve (Figs.\,\ref{fig4}(h) and \ref{fig4}(i)) confirms the value of TC$=3$ and $4$, respectively. The simulation and experimental results show an excellent agreement.

Further, to show an accurate determination of higher order TC of a discrete vortex, we have considered a 1D ring array of N=30 lasers. Using the experimental arrangement shown in Fig.\,\ref{fig2}(a), we have generated discrete vortices with TC$=11$ to 14, and determined their values by averaging their interference patterns with TC$=0$. The results are shown in Fig.\,\ref{fig5}. Figures\,\ref{fig5}(a)-\ref{fig5}(d) and \ref{fig5}(e)-\ref{fig5}(h) show the simulated and experimental results obtained by averaging the interference patterns of TC=11, 12, 13 and 14 with TC=0 (TC=0+TC=11, TC=0+TC=12, TC=0+TC=13 and TC=0+TC=14), respectively. The white dashed circles in Figs.\,\ref{fig5}(a) and \ref{fig5}(e) mark the location of a selected reference laser $j=1$. The location of reference laser remains the same in other averaged interference patterns. As evident, in the averaged interference patterns, corresponding to different TCs, the fringes at the lasers exhibit different behaviour, and enables an accurate identification of corresponding TC of discrete vortices. The analyzed fringe visibility as a function of laser index ($j$) corresponding to TC$=11-14$ is shown in Figs.\,\ref{fig5}(i)-\ref{fig5}(l), respectively. The phase distribution of discrete vortex with different TCs, formed with a 1D ring array of N=30 lasers cane be given as $\phi_{j}=\mathrm{TC}.(2\pi(j-1)/30)$, where $j=1,~2,~3,...30$. The observed variation in the fringe visibility as a function of laser index ($j$) can be explained with the same reasons as explained above for discrete vortex with N=10 lasers. It is clearly evidenced that the averaging of interference patterns TC=0+TC=11, TC=0+TC=12, TC=0+TC=13 and TC=0+TC=14 leads to distinct variations in the fringe visibility as a function of laser index, and produces different number of dips corresponding to different values of TCs. For example, averaged interference pattern of TC=0+TC=11 produces eleven number of dips, which confirms the value of TC=11 (Fig.\,\ref{fig5}(i)). Similarly, the observation of twelve, thirteen and fourteen dips corresponding to TC=0+TC=12, TC=0+TC=13, and TC=0+TC=14, confirms the values of TC$=12-14$, respectively. We have observed an excellent agreement between the numerical and experimental results, which indicates an accurate determination of high order TCs of discrete vortices. 
\begin{figure*}[htbp]
\centering
\includegraphics[width=17.5cm]{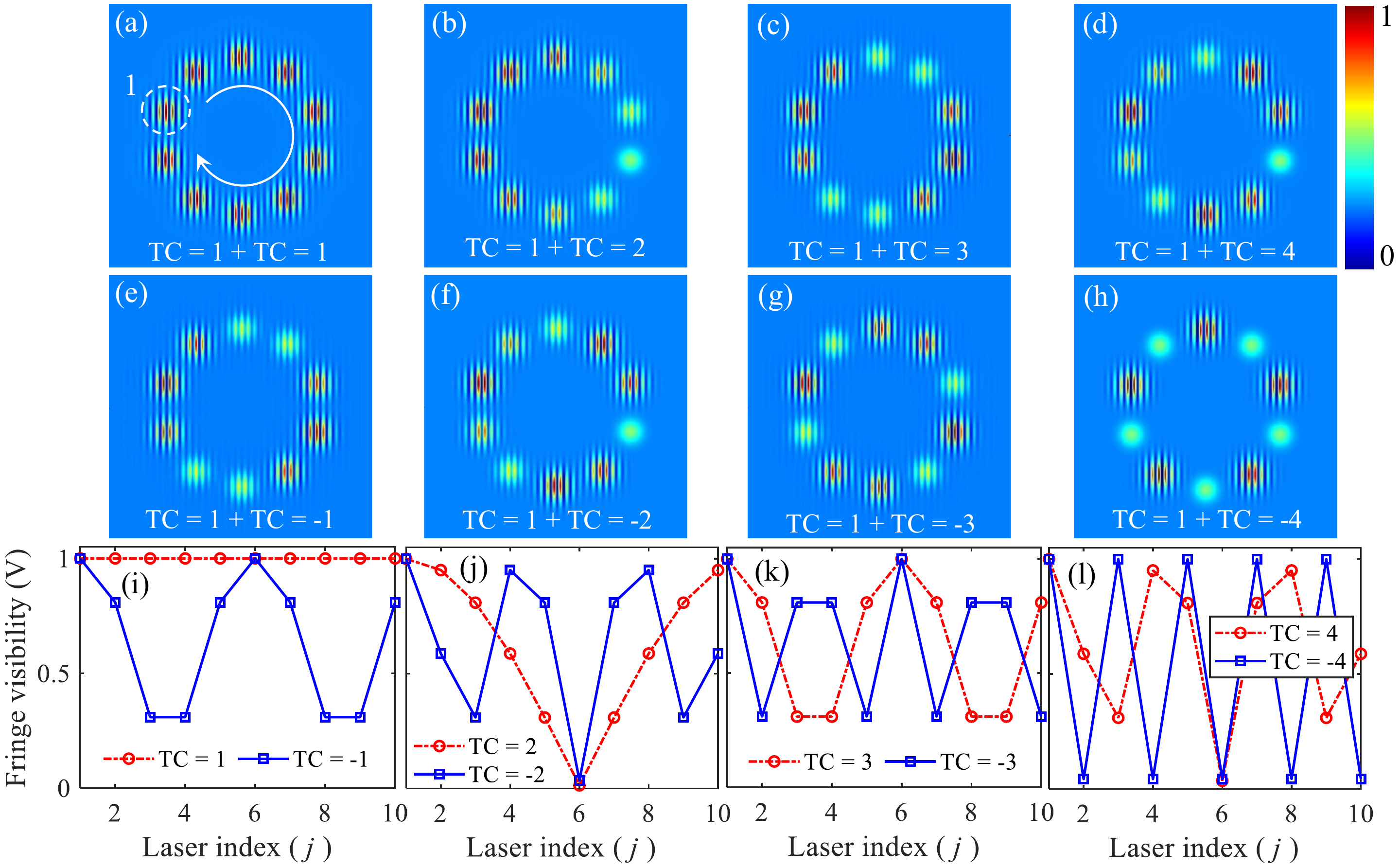}
\caption{\label{fig6} Simulation results for determining the sign of TC of a discrete vortex with N=10 lasers. The averaging of interference patterns of TC$=1$ and (a) TC=$1$, (b) TC$=2$, (c) TC$=3$, and (d) TC$=4$. The averaging of interference patterns of TC$=1$ and (e) TC$=-1$, (f) TC$=-2$, (g) TC$=-3$, and (h) TC$=-4$. (i-l) The fringe visibility as a function of laser index, corresponding to (a,e), (b, f), (c, g), and (d, h), respectively. A dashed white circle in (a) marks the location of a selected reference laser $j=1$. }
\end{figure*}
\begin{figure*}[htbp]
\centering
\includegraphics[width=17.5cm]{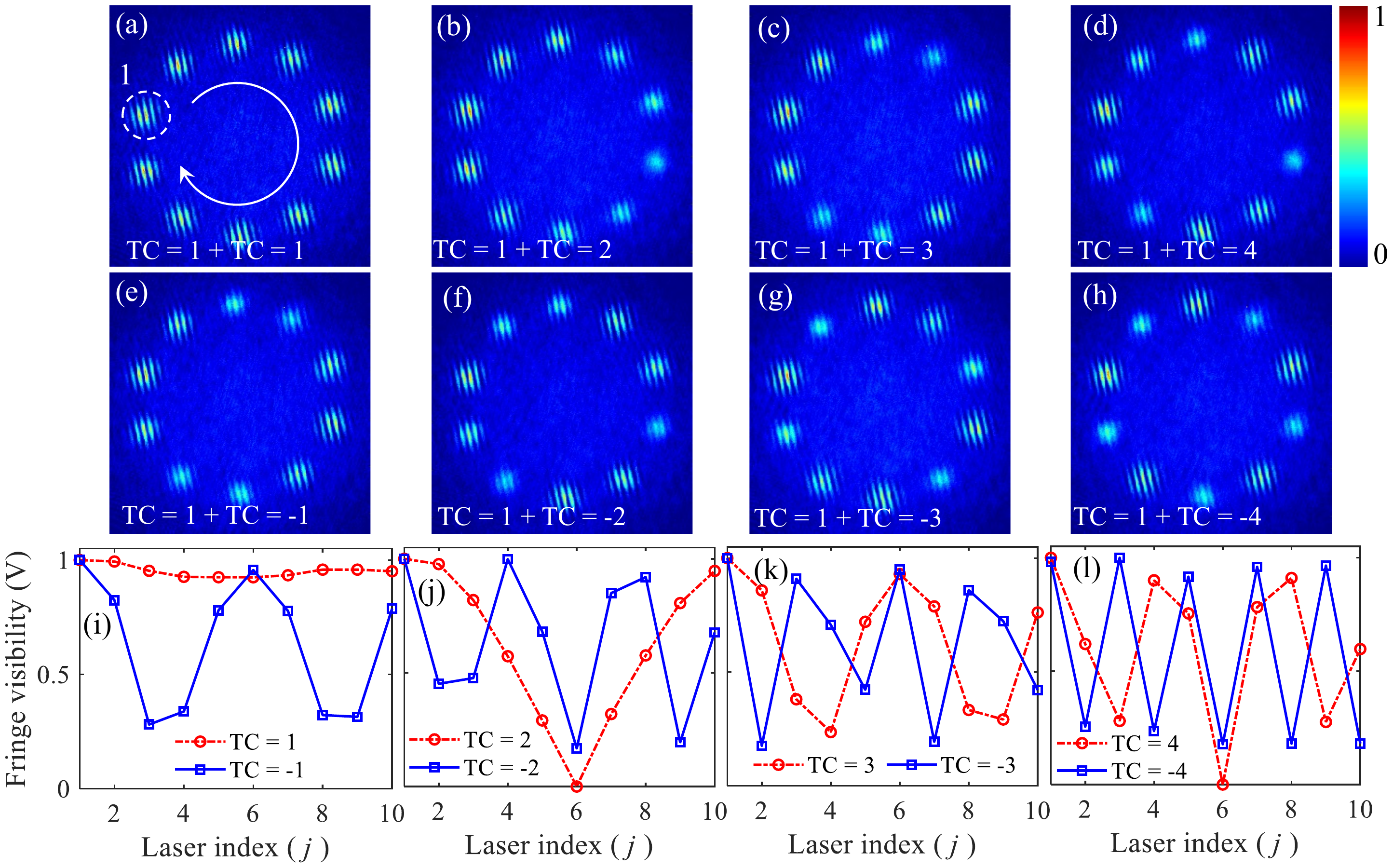}
\caption{\label{fig7} Experimental results for determining the sign of TC of a discrete vortex with N=10 lasers. The averaging of interference patterns of TC$=1$ and (a) TC=$1$, (b) TC$=2$, (c) TC$=3$, and (d) TC$=4$. The averaging of interference patterns of TC$=1$ and (e) TC$=-1$, (f) TC$=-2$, (g) TC$=-3$, and (h) TC$=-4$. (i-l) The fringe visibility as a function of laser index, corresponding to (a,e), (b, f), (c, g), and (d, h), respectively. A dashed white circle in (a) marks the location of a selected reference laser $j=1$. }
\end{figure*}

So far, we have successfully shown an accurate determination of maginutude of TC of a discrete vortex by averaging the interference patterns of TC$\neq 0$ with TC$=0$. However, with this averaging the same results are obtained for both positive and negative values of TC, and can not distinguish the sign of TC$\neq 0$. To determine the sign of TC of a discrete vortex, we average the interference pattern of unknown discrete vortex having TC$\neq0$ with the interference pattern of known discrete vortex having TC$=1$ (instead of TC$=0$ as earlier). To show it, we have considered discrete vortices with different TCs in a 1D ring array of N=10 lasers, as shown in Fig.\,\ref{fig6}.

Figures\,\ref{fig6}(a)-\ref{fig6}(d) show the simulated averaged interference patterns of positive TC=1, 2, 3 and 4 with TC=1 (TC$=1+$TC$=1$, TC$=1+$TC$=2$, TC$=1+$TC$=3$ and TC$=1+$TC$=4$), respectively. Figures\,\ref{fig6}(e)-\ref{fig6}(h) show the simulated averaged interference patterns of negative TC$=-1,~-2,~-3,$ and $-4$ with TC=1 (TC$=1+$TC$=-1$, TC$=1+$TC$=-2$, TC$=1+$TC$=-3$ and TC$=1+$TC$=-4$, respectively. A white dashed circle in Fig.\,\ref{fig6}(a) represents the location of a selected reference laser $j=1$, which also remains the same in all other interference patterns (Figs.\,\ref{fig6}(b)-\ref{fig6}(h)). As evident, in the averaged interference patterns of positive and negative TCs with TC=1, the fringes at the lasers appear differently, thus enabling to distinguish the sign of TC. These different interference patterns are again anticipated by the different phase distribution of discrete vortices having different TCs. The analyzed fringe visibility as a function of laser index ($j$), corresponding to Figs.\,\ref{fig6}(a)-\ref{fig6}(h), is shown in Figs.\,\ref{fig6}(i)-\ref{fig6}(l). The red dashed curve with circles shows the visibility curve when interference patterns of positively charged TC$=+1$ to $+4$ are averaged with TC$=+1$, whereas, solid blue curve with squares denote the case when interference patterns of negatively charged TC$=-1$ to $-4$ are averaged with TC$=+1$. As evident, the variation of fringe visibility with laser index ($j$) for the positive and negative TCs is found to be different, and thus enables to identify the sign of TC. In particular, for TC$=+1$, the fringe visibility is found to be 1 for all the lasers (no variation), and no dip is observed (dashed red curve with circles in Fig.\,\ref{fig6}(i)). Whereas, for TC$=-1$, the fringe visibility shows variation, and two dips are observed (solid blue curve with squares in Fig.\,\ref{fig6}(i)). 

As discussed earlier in Figs.\,\ref{fig3}-\ref{fig5}, when the interference patterns of positive and negative TCs are averaged with the interference pattern of TC$=0$, the number of dips in the fringe visibility curve is found to be proportional to the magnitude of TC. For example, for TC$=\pm 1$ only a single dip is observed in the fringe visibility curve (Fig.\,\ref{fig3}(j)). However, when the interference patterns of positive and negative TCs are averaged with the interference pattern of TC$=+1$, the number of dips in the fringe visibility curve decreases by one for the positive TCs, and increases by one for the negative TCs. For example, in Fig.\,\ref{fig6}(i), no dip is observed for TC$=+1$ and two dips are observed for TC$=-1$. Similarly, for TC$=+2$, $+3$ and $+4$, the number of dips are observed to be one, two and three, respectively (red dashed curve with circles in Figs.\,\ref{fig6}(j)-\ref{fig6}(l)). Whereas, for TC$=-2$, $-3$ and $-4$, the number of dips are observed to three, four and five, respectively (solid blue curve with squares in Figs.\,\ref{fig6}(j)-\ref{fig6}(l)). This clearly indicates that the observation of decrease/increase in the number of dips in the fringe visibility curve provides an accurate information of positive/negative sign of TC of a discrete vortex. 

We have also verified these findings experimentally, the results are shown in Fig.\,\ref{fig7}. The experimental results show excellent agreement with the simulations. Thus, averaging the interference patterns of positive and negative TCs with the interference pattern of TC$=+1$ provides an accurate information of the sign of TC of a discrete vortex. This approach can be used for identifying the sign of TC of any order.

\begin{figure*}[htbp]
\centering
\includegraphics[width=16.5cm]{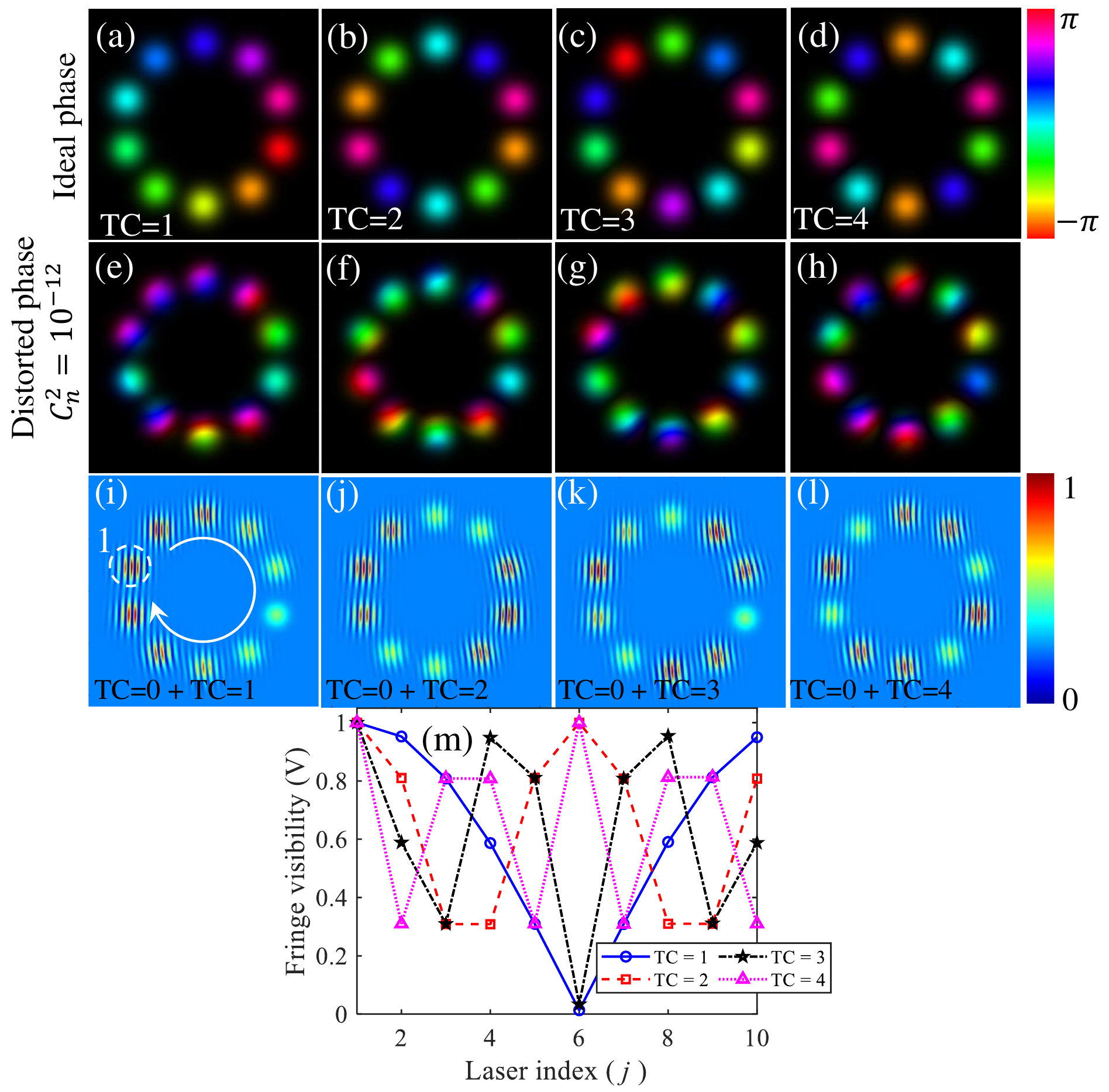}
\caption{\label{fig8} Simulation results showing the effect of phase disorder on the determination of topological charge of a discrete vortex with N=10 lasers. The discrete vortex with topological charges TC=1 to 4 having (a-d) ideal phase distribution, and (e-h) disordered phase distribution. The average of interference patterns of TC$=0$ and (i) TC$=1$, (j) TC$=2$, (k) TC$=3$, and (l) TC$=4$. (m) The fringe visibility as a function of laser index corresponding to (i-j). A white dashed circle in (a) marks the location of a selected reference laser.}
\end{figure*}

Next, we have checked the robustness of our method against the phase disorder (for example, aberrations due to misalignment of optical components). We have generated phase disorder in a range [-$\pi$ to $\pi$] through a random phase screen using Monte Carlo method, that behaves like a disorder media of desired length and having disorder strength $C^{2}_{n}$ \cite{Dev:21-1}. We have considered a phase disorder with a strong strength $C^{2}_{n}=10^{-12}$. The simulated results are shown in Fig.\,\ref{fig8}. Figures\,\ref{fig8}(a)-\ref{fig8}(d) show the ideal phase distribution of discrete vortices with TC=1 to 4 in a 1D ring array of N=10 lasers. After multiplying the phase disorder, the distorted phase distributions of discrete vortices with TC=1 to 4 are shown in Figs.\,\ref{fig8}(e)-\ref{fig8}(l). It is clearly evident that most of the lasers in the array consist of more than a single phase.

To show whether our method can still determine accurately the information of TC, we have measured the interference patterns of 1D ring array of lasers with TC=1 to 4 and TC$=0$ under the same phase disorder, using Fig.\,\ref{fig2}(a). The averaged interference patterns of TC=1 to 4 with TC=0 are shown in Fig. \ref{fig8}(i)-\ref{fig8}(l). It is evident that in the averaged interference patterns of different TCs, the fringes at the lasers are distributed differently with different fringe visibility. Further, the orientations of the fringes at the lasers are found to be different (unlike ideal discrete vortices in Figs.\,\ref{fig2} and \ref{fig3}). This is anticipated by the multiple phase structure in each laser. The analyzed fringe visibility as a function of laser index ($j$) for the all TC=1 to 4 (blue solid curve with circle for TC$=1$, red dashed curve with squares for TC$=2$, black dot-dashed curve with star for TC$=3$ and pink dotted curve with triangles for TC$=4$), are shown in Fig.\,\ref{fig8}(m). As evident, for different values of TC, the variation in the fringe visibility consists of different number of dips. In particular, corresponding to TC=1 to 4, the number of dips in the fringe visibility are found to be one, two, three and four, respectively, which clearly identifies the magnitude of TCs.
\begin{figure}[htbp]
\centering
\includegraphics[width=8.0cm]{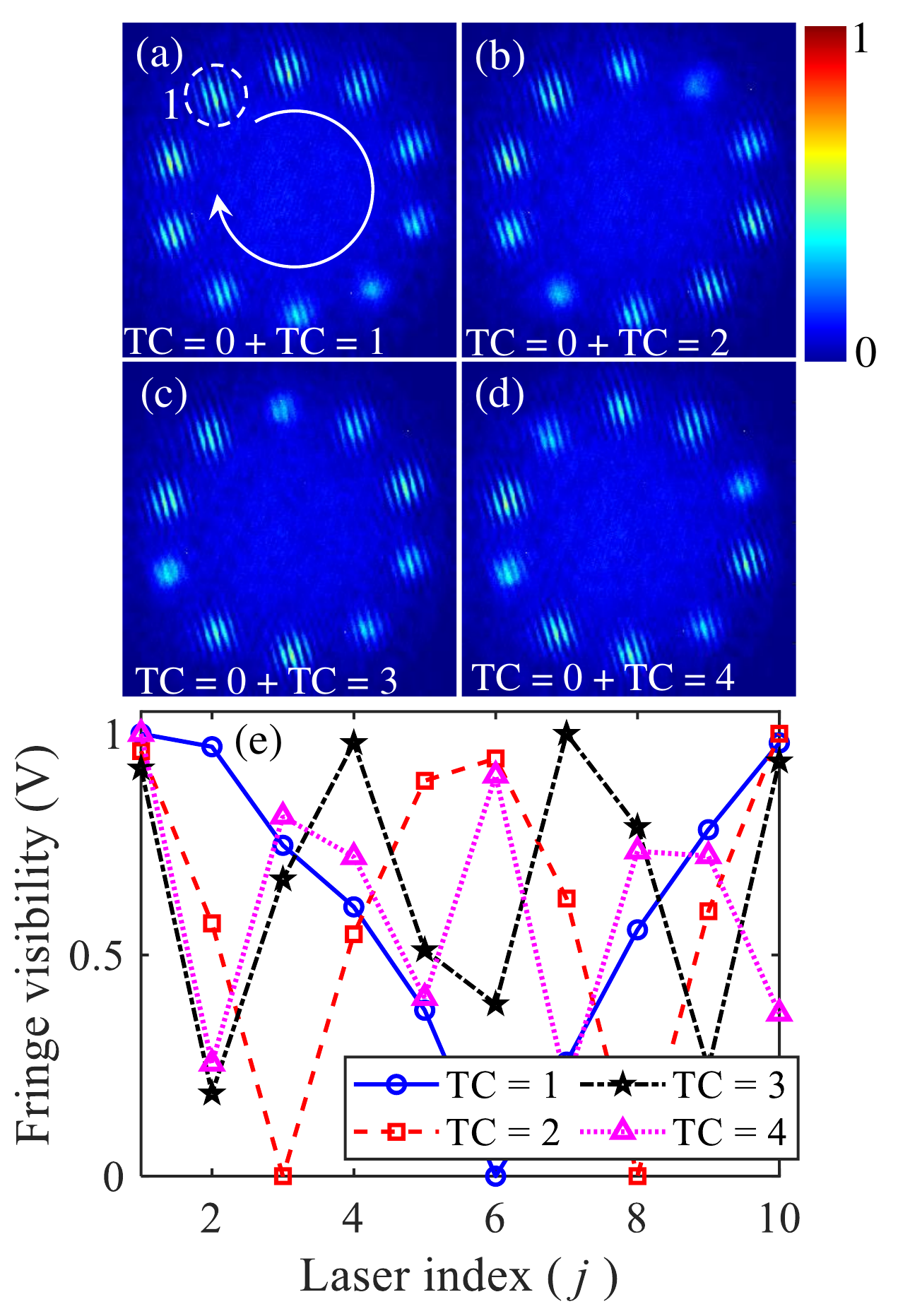}
\caption{\label{fig9} Experimental results showing the effect of phase disorder on the determination of topological charge of a discrete vortex with N=10 lasers. The average of interference patterns of TC$=0$ and (a) TC$=1$, (b) TC$=2$, (c) TC$=3$, and (d) TC$=4$. (m) The fringe visibility as a function of laser  index corresponding to (a-d). Note, the phase disorder is kept the same as in Fig.\,\ref{fig8}. A white dashed circle in (a) marks the location of a selected reference laser.}
\end{figure}

We have also verified experimentally the effect of phase disorder on the determination of TC. The results for different values of TC=1 to 4, under the same phase disorder strength $C^ {2}_{n}=10^{-12}$ are presented in Fig.\,\ref{fig9}. Experimentally, the phase disorder on the discrete vortex is realized by imposing random phase on hologram along with the phase distribution of discrete vortex \cite{Dev:21-1}. Figures\,\ref{fig9}(a)-\ref{fig9}(d) show the averaged interference patterns of TC= 1 to 4 with TC=0 (TC=0+TC=1, TC=0+TC=2, TC=0+TC=3 and TC=0+TC=4), respectively, indicating the different fringe distribution at the lasers in each case. Corresponding to Figs.\,\ref{fig9}(a)-\ref{fig9}(d), the analyzed fringe visibility as a function of laser number ($j$) is shown in Fig.\,\ref{fig9}(e). As evidenced, corresponding to different values of TC=1 to 4, the number of dips in the fringe visibility variation is found to be one (blue solid curve with circles), two (red dashed curve with squares), three (black dot-dashed curve with stars) and four (pink dotted curve with triangles), respectively. This clearly depicts that our approach is highly robust against the phase disorder, as it does not affect the accurate determination of magnitude of TC of a discrete vortex. Similarly, the sign of these disordered TCs can be determined by averaging their interference patterns with the interference pattern of disordered TC=1.

The numerical and experimental results show an excellent agreement, and clearly depicts that our method accurately determines the magnitude and sign of TC of a discrete vortex. The determination of TC is not affected by the imperfections such as the phase disorder caused by aberrations in the system. Further, the method can be used for determining any small to large values of TC.

\section{Conclusions} \label{concl}
We have presented a new method for probing the magnitude and sign of an unknown TC of a discrete vortex, which is formed by an array of lasers in 1D ring geometry. The method relies on measuring the interference pattern of a discrete vortex, which is obtained by interfering a single selected reference laser with itself and with all the other lasers in a 1D ring array. The discrete laser arrays with TC$=0$ and TC$\neq0$ have different phase distributions, thus produce interference patterns with shifted fringes. The averaging of these phase shifted interference patterns give rise to a variation in the fringe visibility as a function of laser number $(j)$ in the array, thus enables identification of TC. The magnitude of TC of a discrete vortex is found to be proportional to the number of dips observed in the fringe visibility curve. The sign of TC is determined by averaging the interference patterns of unknown discrete vortices (TC$\neq0$) with known TC$=+1$. The number of dips in the visibility curve decreases by one for a positive TC, and increases by one for a negative TC. Our method works well for any values of TC of a discrete vortex.

Further, we have also verified our method against the phase disorder that may occur due to the presence of aberrations in the system. It is found that phase disorder does not influence an accurate determination of TC of a discrete vortex. We have found an excellent agreement between the experimental results and numerical simulations. Our method can be useful in the applications of discrete vortices.

\begin{acknowledgments}
We acknowledge the funding support from the Indian Institute of Technology (IIT) Ropar (ISIRD Grant No. 9-230/2018/IITRPR/3255), Science and Engineering Research Board (Grant no. CRG/2021/003060). Vasu Dev acknowledges the fellowship support by IIT Ropar.
\end{acknowledgments}

\appendix
\section{Phase hologram for the formation of a discrete vortex in a 1D ring array of lasers}\label{appa}
In order to generate discrete vortex with desired TC, the phase and amplitude of the incident Gaussian beam is modulated by phase only holograms \cite{Arrizon:07}. The complex electric field of discrete vortex can be expressed as
\begin{equation}
    U(x,y)=A(x,y)\exp(i\phi(x,y)),\label{eqsb1}
\end{equation}
where the amplitude $A(x,y)$ and the phase $\phi(x,y)$ take values in the intervals [0,1] and [$-\pi$,$\pi$]. The aim is to encode the complex field $U(x,y)$ by means of a phase transmittance function (phase hologram) to incorporate amplitude variations as phase variations, that is, a function $h(x,y)$ must be given by
\begin{equation}
    h(x,y)=\exp\left[i\psi\left(A,\phi\right)\right], \label{eqsb2}
\end{equation}
where $\psi(A,\phi)$ accounts both amplitude and phase variations. To find  the desired form of phase function $\psi(A,\phi)$, $h(x,y)$ can be expressed as Fourier series in the domain of $\psi$ as \cite{Arrizon:07}
\begin{equation}
h(x,y)=\sum_{p=-\infty}^{\infty} c^{A}_{p}\exp(ip\phi),\label{eqsb3}
\end{equation}
where 
\begin{equation}
    c^{A}_{p}=\frac{1}{2\pi}\int_{-\pi}^{\pi}\exp[i\psi(A,\phi)]\exp(-ip\phi)d\phi. \label{eqsb4}
\end{equation}
The field $E(x,y)$ can be recovered from only the first-order term of Eq.\,(\ref{eqsb4}), provided that the following identity is fulfilled
\begin{equation}
    c^{A}_{1}=Ac, \label{eqsb5}
\end{equation}
where $c$ is a positive constant. Further, the phase function $\psi(A,\phi)$ with odd symmetry can be expressed as 
\begin{equation}
    \psi(A,\phi)=f(A)\sin(\phi). \label{eqsb6}
\end{equation}
In this case, the phase transmittance function (Eq.\,\ref{eqsb2}) becomes $h(x,y)=\exp[i.f(A)\sin(\phi)]$. Expressing it in the Fourier series using Jacobi-Anger identity, we get
\begin{equation}
\exp[i.f(A)\sin(\phi)]=\sum_{q=-\infty}^{\infty}J_{q}[f(A)]\exp(iq\phi),\label{eqsb7}
\end{equation}
where $J_{q}$ represents Bessel function of $q$-th order. Using Eqs.\,(\ref{eqsb3}), (\ref{eqsb5}) and (\ref{eqsb7}), we get
\begin{equation}
    c^{A}_{1}=J_{1}[f(A)].\label{eqsb8}
\end{equation}
Therefore, from Eqs.\,\ref{eqsb5} and (\ref{eqsb8})
\begin{equation}
    Ac=J_{1}[f(A)].\label{eqsb9}
\end{equation}
The function $f(A)$ can be determined by numerical inversion of the above Eq.\,(\ref{eqsb9}). The maximum value of c for which Eq.\,(\ref{eqsb9}) can be fulfilled is $\sim0.58$, which corresponds to the maximum value of the first-order Bessel function $J_{1}(x)$, which occurs at $x\approx1.84$. This restricts the $f(A)$ in the interval [0,1.84]. The modulated field consists of first and higher orders, thus to separate the first order, a blazed grating is added to the phase of hologram. Thus the resultant phase will have the following form
\begin{figure*}[htbp]
\centering
\includegraphics[width=13.2cm]{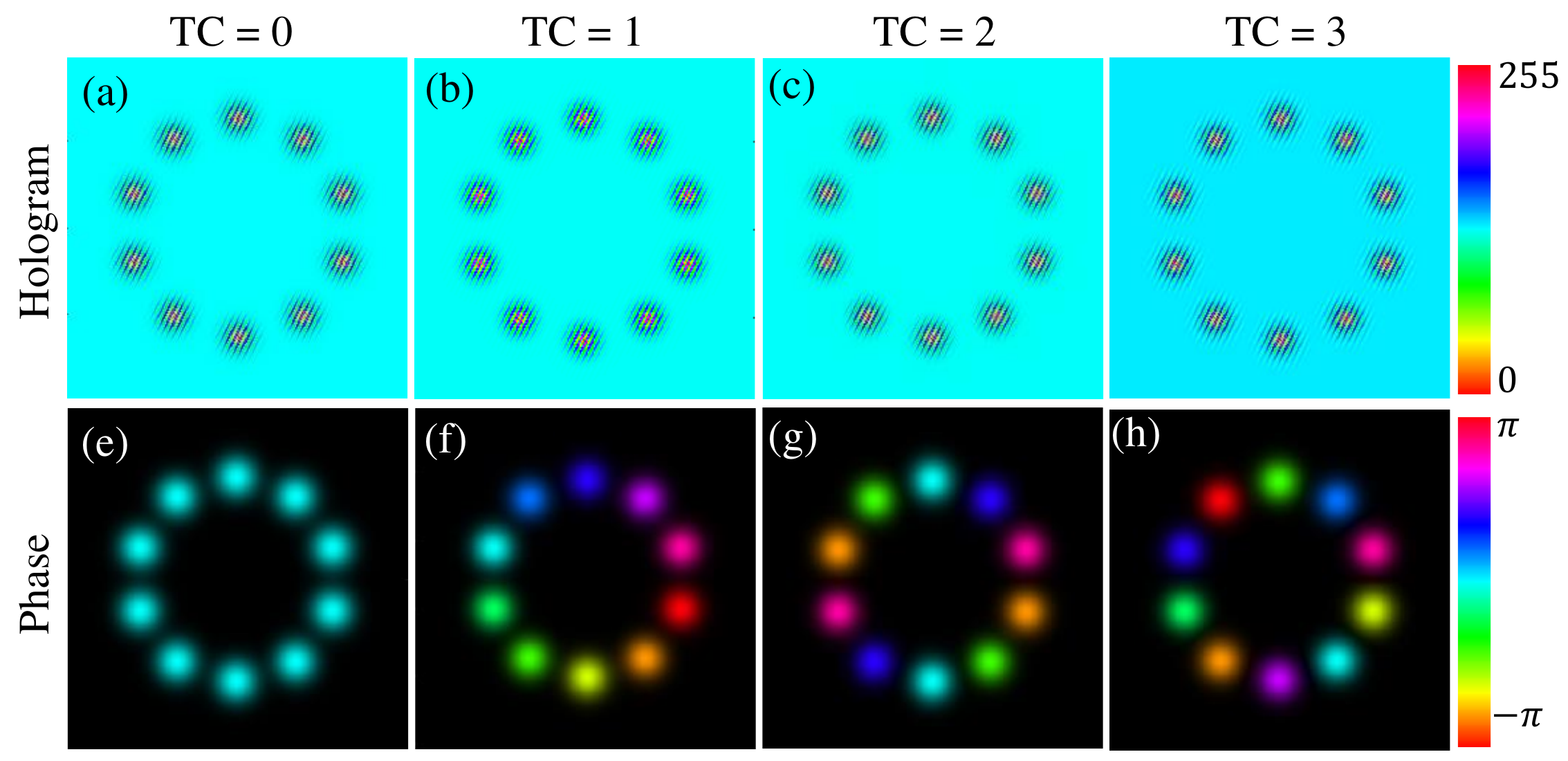}
\caption{\label{fig10}Phase holograms corresponding to (a) TC=0, (b) TC=1, (c) TC=2, and (d) TC=3. The phase distributions of lasers in 1D ring array corresponding to these holograms for (e) TC=0, (f) TC=1, (g) TC=2, and (h) TC=3.}
\end{figure*}
\begin{equation}
    \psi=\psi(A,\phi+2\pi N_{x}x+2\pi N_{y}y),\label{eqsb10}
\end{equation}
where $N_{x}$ and $N_{y}$ denote the grating frequencies along $x$ and $y$ directions, respectively. As per the requirement of SLM, the phase of hologram is divided into 256 levels. The spatial frequencies are chosen as $N_{x}=120$ and $N_{y}=60$. The holograms corresponding to TC$=0$ to 3 and system size N$=10$ are shown in Fig.\,\ref{fig10}(a)-\ref{fig10}(d), respectively. These holograms are used on the SLM to produce the desired phase distributions of lasers in 1D ring array with TC$=0$ to 3, as shown in Fig.\,\ref{fig10}(e)-\ref{fig10}(h), respectively.

The similar procedure for generating discrete vortex from large system size and higher TCs is used. 

\section{Fringe visibility analysis} \label{appb}
\begin{figure*}[htbp]
\centering
\includegraphics[width=18.2cm]{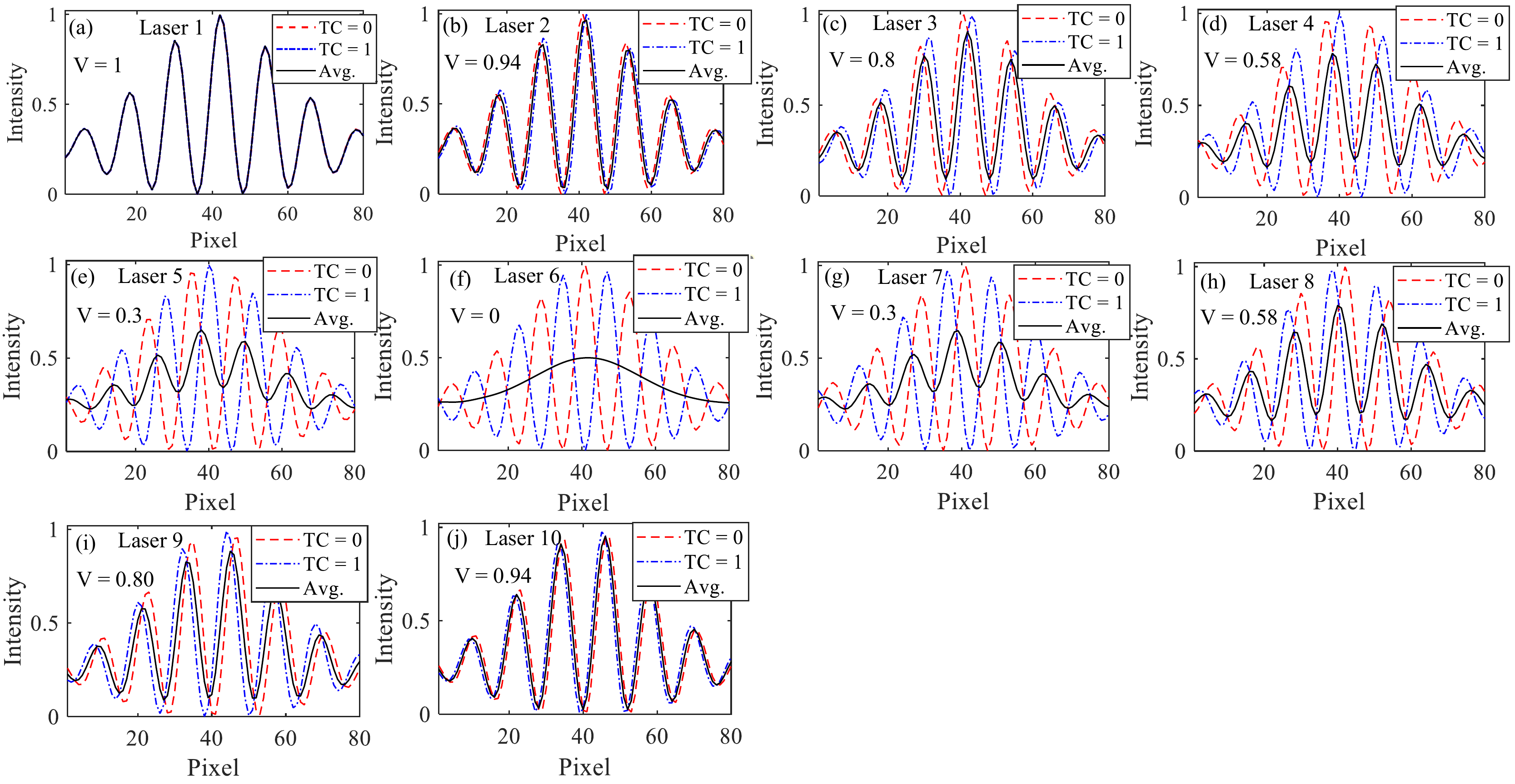}
\caption{\label{fig11} Simulated results. The intensity cross-section in the interference patterns of TC$=0$ , TC$=1$ and averaged interference pattern TC$=0+$TC$=1$ at (a) laser 1, (b) laser 2, (c) laser 3, (d) laser 4, (e) laser 5, (f) laser 6, (g) laser 7, (h) laser 8, (i) laser 9, and (j) laser 10. The blue dot-dashed curve denotes TC$=0$, red dashed curve denotes TC$=1$, and black solid curve denotes TC$=0+$TC$=1$. Note, the intensity cross-sections in TC=0 and TC=1 are normalized to the maximum value of 1.}
\end{figure*}

\begin{figure*}[htbp]
\centering
\includegraphics[width=18.2cm]{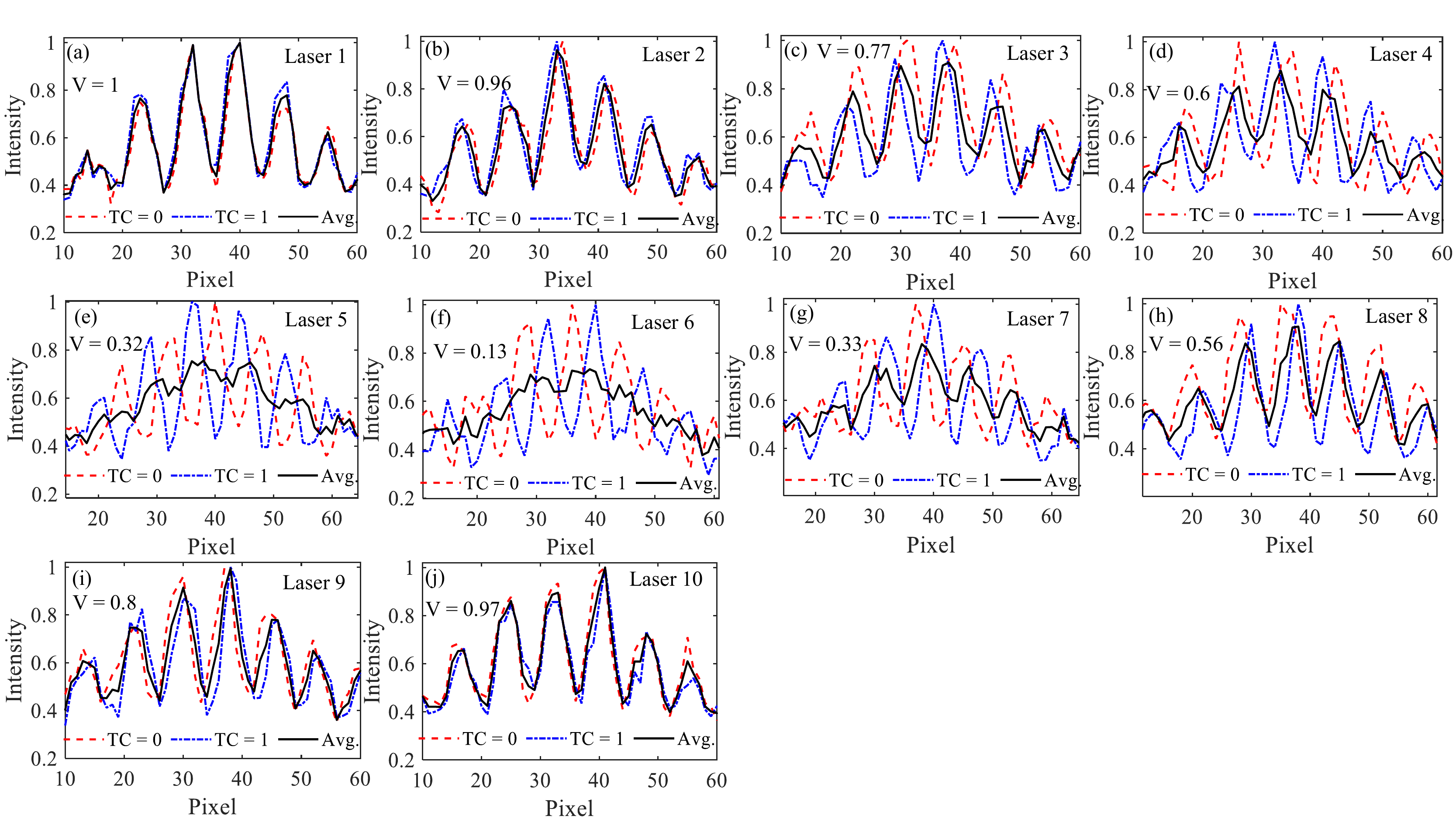}
\caption{\label{fig12}Experimental results with the details same as given in Fig.\,\ref{fig10}.}
\end{figure*}

Here, we have presented the analysis of fringe visibility in the interference patterns obtained for TC=0 (Fig.\,\ref{fig3}(b)) and TC=1 (Fig.\,\ref{fig3}(c)) as well as the averaged interference pattern of these two. Note, the phase distribution of discrete vortex in a ring array of N=10 lasers, is given as $\phi_{j}=\mathrm{TC}.(j-1).(2\pi/10)$, where $j=1,~2,~3...$ denote the laser number in the array. The interference patterns are obtained by using Mach-Zhender interferometer (Fig.\,\ref{fig2}(a)), where a single selected reference laser interferes with itself and with all the other lasers in the array. The results of interference patterns are shown in Figs.\,\ref{fig3}(d)-\ref{fig3}(f) (simulation) and Figs.\,\ref{fig3}(g)-\ref{fig3}(i) (experiment). The analysis of fringe visibility at all lasers in the array is shown in Fig.\,\ref{fig11} (simulation) and Fig.\,\ref{fig12} (experiment). 

The laser $j=1$ is a selected reference laser, where light interferes with itself, and has initial phase difference d$\phi_{1,1}=0$ (Eq.\,(\ref{eq5})) for both TC$=0$ and TC$=1$. The interference expressions at laser 1 in TC=0 and TC=1 can be written as following. Using Eq.\,\ref{eq5})
\begin{equation}
I_{T1}=I_{T2}=2I_{0}\left[1+\cos\left(0.94 k x\right)\right],\label{eqs1}
\end{equation}
The expression of averaged interference pattern is given as
\begin{equation}
    I_{Sum}=\frac{I_{T1}+I_{T2}}{2}=2I_{0}\left[1+\cos\left(0.94 k x\right)\right].\label{eqs2}
\end{equation}

As d$\phi_{1,1}=0$ for both cases of TC=0 and TC=1, so maxima and minima in the fringe pattern (Eq.\,(\ref{eqs1})) occur at the same locations (red dashed curve and blue dot-dashed curve in Figs.\,\ref{fig11}(a) and \ref{fig12}(a)). In both cases, the fringe visibility is found to be $V=1$. The average of these two interference patterns gives no change in the fringe pattern (Eq.\,(\ref{eqs2})), and the fringe visibility remains unchanged $V=1$ (black solid curve in Figs.\,\ref{fig11}(a) and \ref{fig12}(a)). 
\nocite{*}

The interference fringes at laser 2 in TC=0 and TC=1 are obtained when the light from a selected reference laser $j=1$ interferes with the light from laser $j=2$. In TC=0, the initial phase difference between the laser $j=1$ and laser $j=2$ is $d\phi_{1,2}=0$, whereas, in TC=1 it is $d\phi_{1,2}=2\pi/10$. The interference expressions $I_{T1}$ (in TC=0) and $I_{T2}$ (in TC=1) can be given as
\begin{eqnarray}
 I_{T1}&=& 2I_{0}\left[1+\cos\left(0.94 k x\right)\right],\label{eqs3}\\ 
 I_{T2}&=&2I_{0}\left[1+\cos\left(0.94 k x-1.(2\pi/10)\right)\right].\label{eqs4}
\end{eqnarray}
The averaged interference pattern is given as
\begin{equation}
I_{Sum}=I_{0}\left[2+\cos\left(0.94 k x\right)+\cos\left(0.94 k x-1.(2\pi/10)\right)\right].\label{eqs5}
\end{equation}
The interference patterns (Eqs.\,(\ref{eqs3}) and (\ref{eqs4})) consist of fringes with clear maxima and minima, which give rise to the fringe visibility of $V=1$ (red dashed curve and blue dot-dashed curve in Figs.\,\ref{fig11}(b) and \ref{fig12}(b)). As evident, in these interference patterns the fringes (maxima and minima) are shifted relative to each other, which occurs due to the different values of initial phase difference $d\phi_{1,2}$ in TC=0 and TC=1. When these interference patterns with shifted fringes are averaged, the resultant interference pattern (Eq.\,\ref{eqs5}) consists of fringes with reduced fringe visibility $V=0.94$ (black solid curve in Figs.\,\ref{fig11}(b) and \ref{fig12}(b)).

Similarly the interference fringes at other lasers are obtained by interfering the light from a selected reference laser $j=1$ and the light from other lasers $j>2$. In general, in TC$=0$, the phase difference between the selected reference laser $j=1$ and other lasers is given as d$\phi_{1,j\ge1}=0$, whereas, in case of TC$\ne 0$, it is given as d$\phi_{1,j\ge1}=\mathrm{TC}.(j-1).(2\pi/N)$, where $N$ denotes the number of lasers in a 1D ring array. Therefore, in case of TC=0 the interference fringes at lasers $j=3, 4,...10$ occur exactly at the same locations (red dashed curve in Figs.\,\ref{fig11}(c)-\ref{fig11}(j) and Figs.\,\ref{fig12}(c)-\ref{fig12}(j)). However, in case of TC=1 the interference fringes at $j=3, 4,...10$ occur at different locations due to different values of d$\phi_{1,j\ge3}$ (blue dot-dashed curve in Figs.\,\ref{fig11}(c)-\ref{fig11}(j) and Figs.\,\ref{fig12}(c)-\ref{fig12}(j)). Because of this when the interference patterns at the same lasers ($j=3, 4,...10$) for TC=0 and TC=1 are averaged, the resultant interference pattern consists of reduced visibility (black solid curve in Figs.\,\ref{fig11}(c)-\ref{fig11}(j) and Figs.\,\ref{fig12}(c)-\ref{fig12}(j)). For example, at lasers $j=3-6$ the values of reduced fringe visibility are found to be 0.8 (Figs.\,\ref{fig11}(c) and \ref{fig12}(c)), 0.58 (Figs.\,\ref{fig11}(d) and \ref{fig12}(d)), 0.3 (Figs.\,\ref{fig11}(e) and \ref{fig12}(e)), and 0 (Figs.\,\ref{fig11}(f) and \ref{fig12}(f)), respectively. For the lasers $j=7-10$, the values of fringe visibility are found to be 0.3 (Figs.\,\ref{fig11}(g) and \ref{fig12}(g)), 0.58 (Figs.\,\ref{fig11}(h) and \ref{fig12}(h)), 0.80 (Figs.\,\ref{fig11}(i) and \ref{fig12}(i)) and 0.94 (Figs.\,\ref{fig11}(j) and \ref{fig12}(j)), respectively. Note, a small discrepancy between simulated and experimental fringe visibility is anticipated by a small intensity difference between the lasers in the experiment. As evident, the fringe visibility decreases monotonically for lasers $j=1-6$ and after that it again increases from $j=7-10$. The increase in the fringe visibility for $j=7-10$ can be explained as follows. For laser $j=7$, we have d$\phi_{1,7}=1.6.(\frac{2\pi}{10})$, which can also be written as $\left[2\pi-(4.(\frac{2\pi}{10}))\right]$ or $-4.(\frac{2\pi}{10})$. This becomes equivalent to the phase difference between $j=1$ and 5 with the negative sign i.e, -d$\phi_{1,5}$. Note, the positive and negative values of the same magnitude d$\phi$ produce the same shift of fringes in the interference pattern, as shown in Figs.\,\ref{fig1}(a) and \ref{fig1}(c). The phase difference $|d\phi_{1,7}|<|d\phi_{1,6}|$, and due to this the fringe visibility in the averaged interference at laser $j=7$ is found to be higher than laser $j=6$ (and same as for $j=5$). Similarly, the increased values of $V$ at laser $j=8, ~9, ~\mathrm {and} ~10$ can also be explained. 

%\bibliography{apssamp}% Produces the bibliography via BibTeX.
\providecommand{\noopsort}[1]{}\providecommand{\singleletter}[1]{#1}%
\end{document}